\documentclass[aip,jcp,twocolumn,superscriptaddress,showpacs,showkeys,reprint]{revtex4-1}

\usepackage{amssymb}
\usepackage{amsmath}
\usepackage{graphicx}
\usepackage{hyperref}   

\newcommand{\homega}{\hat{\omega}}

\newcommand*{\citen}{}
\DeclareRobustCommand*{\citen}[1]{%
  \begingroup
    \romannumeral-`\x 
    \setcitestyle{numbers}%
    [\cite{#1}]%
  \endgroup
}

\usepackage{epstopdf}

\begin{document}

\title{Sedimentation stacking diagrams of binary mixtures of thick and thin hard rods}

\author{Tara Drwenski}
\email{t.m.drwenski@uu.nl}
\affiliation{Institute for Theoretical Physics, Center for Extreme Matter and Emergent Phenomena, Utrecht University, Princetonplein 5, 3584 CC Utrecht, The Netherlands}

\author{Patrick Hooijer}
\affiliation{Institute for Theoretical Physics, Center for Extreme Matter and Emergent Phenomena, Utrecht University, Princetonplein 5, 3584 CC Utrecht, The Netherlands}

\author{Ren\'{e} van Roij} 
\affiliation{Institute for Theoretical Physics, Center for Extreme Matter and Emergent Phenomena, Utrecht University, Princetonplein 5, 3584 CC Utrecht, The Netherlands}

\date{\today}

\begin{abstract}
    We use Onsager theory and the local density approximation to study sedimentation-diffusion equilibrium density profiles of binary mixtures of thick and thin hard rods. We construct stacking diagrams for three diameter ratios, and find that even a simple spindle-shaped phase diagram with only isotropic-nematic demixing can lead to counter-intuitive stacking sequences such as an isotropic phase sandwiched between two nematic phases. For the most complex phase diagram considered here, we find sixteen distinct stacking sequences, including several with five sedimented layers. By adding sedimentation paths to composition-pressure and density-density phase diagrams and calculating density and composition profiles, we show that conclusions about bulk phase diagrams of binary mixtures on the basis of sedimentation-diffusion equilibria should be drawn warily.
\end{abstract}


\maketitle





\section{Introduction}\label{sect:intro}

Colloidal sedimentation-diffusion (SD) equilibria are the result of a competition between gravitational energy (which favors a high mass density at the bottom) and entropy (which favors a homogeneous distribution of matter). In a sufficiently dilute one-component system, where the particles can be considered non-interacting, the height-dependent density profile is simply the barometric distribution, a fact which can be used to determine the buoyant masses of colloids. Historically, however, Perrin used colloids with a known buoyant mass to determine Boltzmann's constant $k_B$ from the barometric profile.\cite{Perrin1910} For one-component colloidal systems, a single density profile can be used to determine the osmotic equation of state. Therefore sedimentation is an important tool to gain information about thermodynamic properties.\cite{Biben1993, Piazza1993, Rutgers1996,Savenko2004} 

Ultracentrifugation can also be used as a method to study SD equilibrium. Here the gravitational field is replaced by a centrifugal one, which allows for alteration of the (effective) buoyant mass by varying angular frequency of the centrifuge's rotation.\cite{Svedberg1940,Rasa2004,Planken2010}

Gravity can also lead to non-barometric profiles and unexpected new phenomena, for instance when considering colloidal systems with electrostatic interactions. A well-studied example is the SD equilibrium of charged spheres at low salt, where interactions lead to the creation of a macroscopic electric field.\cite{vanRoij2003,Rasa2004,Rasa2005,Zwanikken2005} A microscopic theory arising from a generalization of Archimedes' principle, which accounts for density perturbations in the solvent due to interactions with colloids, was successful in describing experimental results such as denser particles floating on top of a lighter fluid.\cite{Piazza2012,Piazza2013} Other counter-intuitive behavior occurs when considering mixtures, for example, a liquid floating between two gases\cite{Schmidt2004} or a nematic phase floating on top of an isotropic phase in a platelet-sphere mixture.\cite{deLasHeras2012} For a recent review on sedimentation see Ref.~\citen{Piazza2014}.

It is more difficult, however, to draw conclusions about bulk phase diagrams of binary mixtures from SD experiments. Recently, Refs.~\citen{deLasHeras2013,deLasHeras2015} showed that the chemical potential representation of a bulk phase diagram together with the local density approximation can be used to create a stacking diagram, which gives all possible stacking sequences of a binary mixture in SD equilibrium. In fact, an application of this method to patchy colloidal mixtures in SD equilibrium showed a good agreement between this method and Monte Carlo simulations which also included lateral walls.\cite{deLasHeras2016} In order to apply this method, it is necessary to revisit established results on binary mixtures in order to obtain bulk phase diagrams in the plane of chemical potentials.

One of the theoretically most accessible model systems of binary hard-core mixtures is rod-rod mixtures, since Onsager theory\cite{Onsager} can be used to obtain accurate results in the needle limit. There is a wealth of work concerning long and short rod mixtures, which have a rich phase behavior including isotropic-nematic demixing with strong fractionation\cite{Onsager,Flory1978,Lekkerkerker1984,Odijk1985} and also, when the length ratio is sufficiently high, nematic-nematic demixing.\cite{Flory1978,Birshtein1988,Vroege1993,vanRoij1996b,Hemmer1999,Varga2000,Speranza2002} Thick and thin rod mixtures can also exhibit isotropic-isotropic demixing, in addition to isotropic-nematic and nematic-nematic phase separation.\cite{Sear1995,vanRoij1996,vanRoij1998,Hemmer2000} These mixtures were also studied theoretically for finite aspect ratio rods\cite{Varga2003,Varga2005} and in experiments,\cite{Purdy2005} and are therefore a well-understood binary system.

Here we apply the method of Refs.~\citen{deLasHeras2013,deLasHeras2015} to a binary mixture of thick and thin needles. First, we review Onsager theory for a binary mixture and display phase diagrams for three diameter ratios. We then construct the stacking diagrams using the chemical potential representation of the phase diagrams. In addition, we show example sedimentation paths, which are given by lines in the plane of chemical potentials, but which we also translate into contours in the composition-pressure and density-density representations. We conclude by discussing relevant experimental results and possible difficulties in determining bulk properties from sedimentation profiles.

\section{Onsager theory for binary mixtures}\label{sect:theory}

We briefly review Onsager theory extended to binary mixtures of thick and thin rods, closely following Ref.~\citen{vanRoij1998}. We consider $N_\sigma$ hard rods of two species $\sigma=1,2$ with equal lengths $L$ but different diameters $D_1$ and $D_2$, suspended in a solvent with volume $V$ at room temperature $T$. The two species of rods have a diameter ratio $d=D_2/D_1 >1$ and the total number of rods is $N=N_1+N_2$. We assume needle-shaped rods with $L \gg D_1,D_2$ such that the excluded volume for a pair of rods of species $\sigma$ and $\sigma'$ is $(D_\sigma + D_{\sigma'})L^2 |\sin \gamma|$, with $\gamma$ the angle between the two rods. We define the dimensionless number density as $c= b N/V$, with $b=(\pi/4)L^2 D_1$ the second virial coefficient of the thin rods in the isotropic phase, and the composition fraction of thick rods as $x=N_2/N$. Each species has an orientation distribution function $\psi_\sigma(\homega)$, where $\homega$ is the orientation of the long axis of the rod. The free energy $F$ is a functional of $\psi_1(\homega)$ and $\psi_2(\homega)$ given by\cite{Onsager, Lekkerkerker1984, vanRoij1998}
    \begin{equation}\label{eq:freeEnergy}
        \frac{\beta F[\psi_1, \psi_2]}{N}=f[\psi_1, \psi_2]=\log c + f_\text{mix}+f_\text{or}+f_{ex},
    \end{equation}
where the mixing contribution $f_\text{mix}$, the orientation contribution $f_\text{or}$, and the excess contribution $f_\text{ex}$ due to excluded volume interactions are given by
    \begin{align}
        f_\text{mix} &= x \log x +(1-x) \log(1-x), \nonumber \\
        f_\text{or} &= \int \, d\homega \left[ (1-x)\psi_1(\homega) \log\psi_1(\homega) + x \psi_2(\homega) \log\psi_2(\homega) \right], \nonumber\\
        f_\text{ex} &= \frac{4c}{\pi} \iint \, d\homega\, d\homega' |\sin \gamma|\, \left[ (1-x)^2\psi_1(\homega) \psi_1(\homega') \right. \nonumber\\
        & \qquad \left.+ x(1-x)(1+d)\psi_1(\homega) \psi_2(\homega') +x^2d\, \psi_2(\homega) \psi_2(\homega') \right].\nonumber
    \end{align}

For a given $c$ and $x$, the equilibrium orientation distribution functions can be obtained by minimizing $f[\psi_1, \psi_2; c,x]$ with respect to $\psi_\sigma(\homega)$ at fixed normalizations $\int \psi_\sigma(\homega) \, d\homega = 1$, which gives the equilibrium distributions as solutions of the integral equations\cite{Vroege1992,vanRoij1998}
    \begin{align}
        C_1 &= \log \psi_1(\homega) + \frac{4c}{\pi} \int \, d\homega |\sin \gamma| \nonumber \\
                & \quad \times \left[ 2(1-x)\psi_1(\homega') + x(1+d)\psi_2(\homega') \right], \label{eq:EL1}\\
        C_2 &= \log \psi_2(\homega) + \frac{4c}{\pi} \int \, d\homega |\sin \gamma| \nonumber\\
                & \quad \times \left[ (1-x)(1+d)\psi_1(\homega') + 2xd\, \psi_2(\homega') \right], \label{eq:EL2}
    \end{align}
where $C_\sigma$ are Lagrange multipliers that ensure proper normalizations of $\psi_\sigma(\homega)$.

At low enough $c$, the isotropic distributions $\psi_{\sigma}^\text{I} = 1/(4\pi)$ are the only (stable) solutions to Eqs.~\eqref{eq:EL1} and \eqref{eq:EL2},\cite{Kayser1978,Mulder1989} which gives for the free energy of the isotropic phase
    \begin{equation}
        \label{eq:freeEnergyIso}
        f^\text{I}(x,c) = \log c +f_\text{mix} - \log(4\pi)+ c (1+(d-1)x).
    \end{equation}

\begin{figure}[tb]
        \centering
                \includegraphics[width=\linewidth]{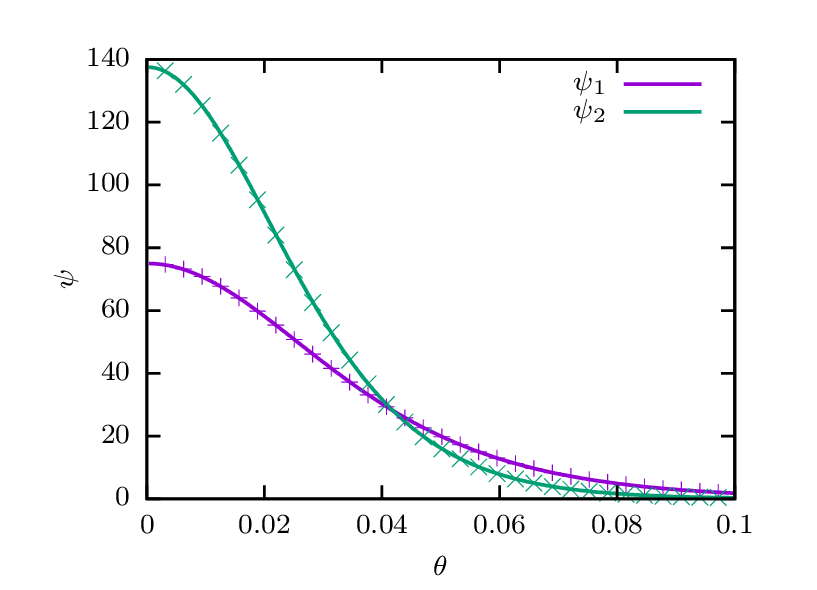}
            \caption{Orientation distribution functions $\psi_1$ and $\psi_2$ of thin and thick needles as a function of the polar angle $\theta$ (in units of radians) for diameter ratio $d=4.2$, composition $x=0.99$, and density $c=8$ for a grid of 500 angles (symbols) and of 5000 angles (lines). The order parameters for the thin and thick rods are $S_1 = 0.996$ and $S_2=0.998$, respectively.}\label{fig:psi}
    \end{figure}

At higher densities, the excluded volume becomes more important in Eqs.~\eqref{eq:EL1} and \eqref{eq:EL2} and the minimizing orientation distribution functions become peaked around a nematic director $\hat{n}$. If we choose a coordinate system with the $z$-axis parallel to $\hat{n} = (0,0,1)$, the unit vector $\homega$ can be written as $\homega=(\sin \theta \, \cos \varphi, \sin \theta \, \sin \varphi, \cos\theta)$, with $\varphi$ the azimuthal angle and $\theta$ the polar angle with respect to $\hat{n}$. In the uniaxial nematic states of interest here, the orientation distribution function is independent of the azimuthal angle $\varphi$ and has up-down symmetry, and so we can write $\psi(\homega)=\psi(\theta) = \psi(\pi - \theta)$. To determine the orientation distribution function for the nematic phase, we solve Eqs.~\eqref{eq:EL1} and \eqref{eq:EL2} using a iterative scheme on a discrete grid of polar angles $\theta \in [0,\pi/2)$.\cite{Herzfeld1984, vanRoij2005} In Ref.~\citen{vanRoij1998} a scaling analysis was used to determine the extremely peaked orientation distribution functions in the high-density limit. However, with present-day computational resources such a scaling is not needed in the density regime of interest here. This is illustrated in Fig.~\ref{fig:psi}, where the orientation distribution functions $\psi_1(\theta)$ and $\psi_2(\theta)$ for both 500 and 5000 $\theta$-angles are in good agreement in a region of high ordering where the nematic order parameters $S_\sigma = \langle (3 \cos^2 \theta -1)/2 \rangle_\sigma$ are $S_1 = 0.996$ and $S_2=0.998$. Therefore we simply employ a $\theta$-grid of 500 angles, which we deem sufficiently accurate for the present purposes. For a given $d$, $x$, and $c$, we numerically calculate the nematic equilibrium orientation distribution functions $\psi_1(\theta)$ and $\psi_2(\theta)$ and insert them into Eq.~\eqref{eq:freeEnergy}, which yields the equilibrium free energy of the nematic phase, $f^\text{N}(x,c)$. 

The pressure $P$ (in units of $k_B T/b$) and chemical potentials $\mu_\sigma$ (in units of $k_B T$) can be found from the free energy, using
    \begin{align}
       & \beta b P^\alpha  = c^2\left( \frac{\partial f^\alpha(x,c) }{\partial c} \right)_{x},\nonumber\\
        &\beta \mu_1^\alpha = f^\alpha(x,c) +c \left( \frac{\partial f^\alpha(x,c) }{\partial c} \right)_{x} + (1-x)\left( \frac{\partial f^\alpha(x,c) }{\partial x} \right)_{c} ,\nonumber\\
        &\beta \mu_2^\alpha = f^\alpha(x,c) +c \left( \frac{\partial f^\alpha(x,c) }{\partial c} \right)_{x} -x\left( \frac{\partial f^\alpha(x,c) }{\partial x} \right)_{c},
    \end{align}
where $\alpha =\text{I,N}$ denotes the isotropic or nematic phase, respectively.

Coexistence between two states ($x^\alpha$, $c^\alpha$) and ($x^{\alpha^\prime}$, $c^{\alpha^\prime}$) is found using the conditions of chemical and mechanical equilibrium: $\mu_1^\alpha = \mu_1^{\alpha^\prime}$, $\mu_2^\alpha = \mu_2^{\alpha^\prime}$, and $P^\alpha = P^{\alpha^\prime} $.
 
As shown in Ref.~\citen{vanRoij1998}, there are four distinct phase diagram topologies for all possible diameter ratios $d>1$. For all diameter ratios there is isotropic-nematic demixing and for $d<4$ this is the only demixing. For $4 \leq d <4.29$ there is also nematic-nematic demixing ending in a consolute point, while for $d \geq 4.29$ this nematic-nematic demixing is no longer closed by the consolute point. For $d\geq 8$ there is an additional isotropic-isotropic demixing. Here we will examine three diameter ratios ($d=3.5$, $4.2$, $10$) which have increasing complexity and show all possible features.

\begin{figure*}[htbp]
        \centering
                \includegraphics[width=\linewidth]{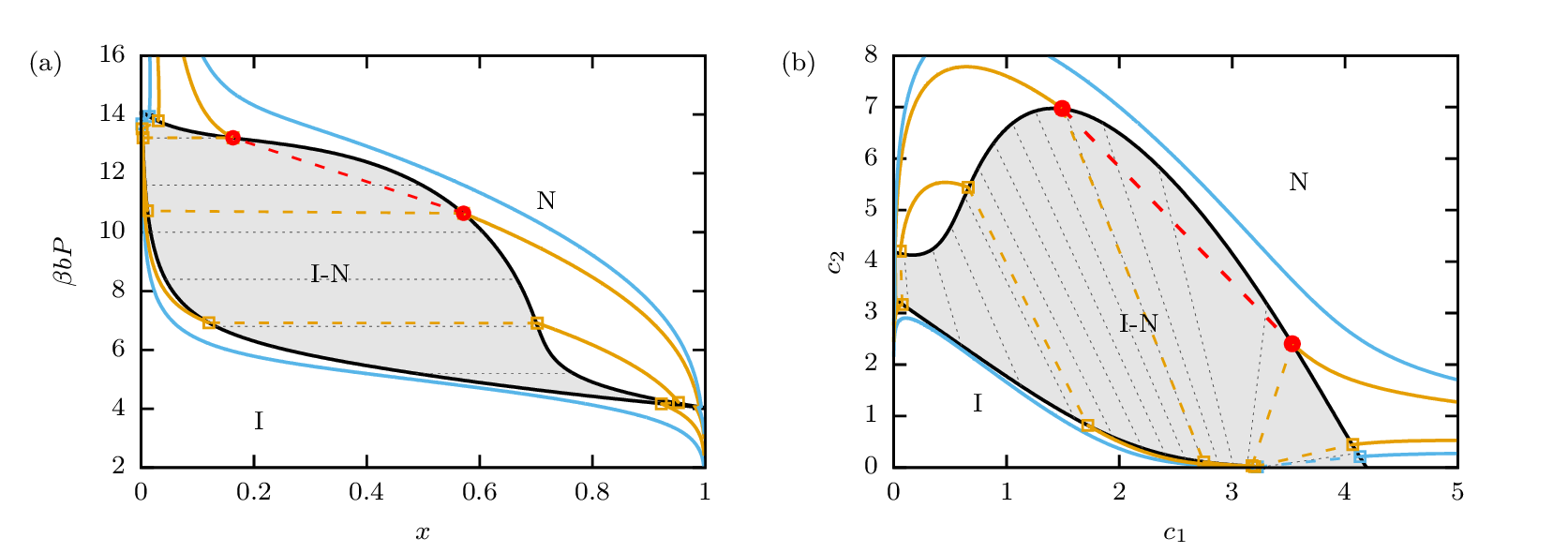}
            \caption{Bulk phase diagram for diameter ratio $d=3.5$ in (a) the composition $x$ - pressure $P$ and (b) the density $c_1$- density $c_2$ representations, where $x$ denotes the mole fraction of thicker rods (species 2). The tie-lines connecting coexisting isotropic (I) and nematic (N) phases are shown by dashed lines. The colored curves indicate sedimentation paths discussed in Section \ref{sect:sedimentation}.}
            \label{fig:d3p5}
     \end{figure*}

For the diameter ratio $d=3.5$, we show in Fig.~\ref{fig:d3p5} the phase diagram in (a) the composition $x$-pressure $P$ and (b) density $c_1$-density $c_2$ representations, where we define the dimensionless number density of thin rods as $c_1 = c(1-x)$ and the density of thick rods as $c_2 = cxd$. For this diameter ratio, the only stable coexistence is found between isotropic (I) and nematic (N) phases.\cite{vanRoij1998} The spindle-shaped I-N coexistence region (shaded with tie-lines shown as dashed lines) shows considerable fractionation, since the thicker rods have a stronger tendency to orientational order. Fig.~\ref{fig:d3p5}(a) is in quantitative agreement with previous results from Ref.~\citen{vanRoij1998}.

\begin{figure*}[htbp]
        \centering
                \includegraphics[width=\linewidth]{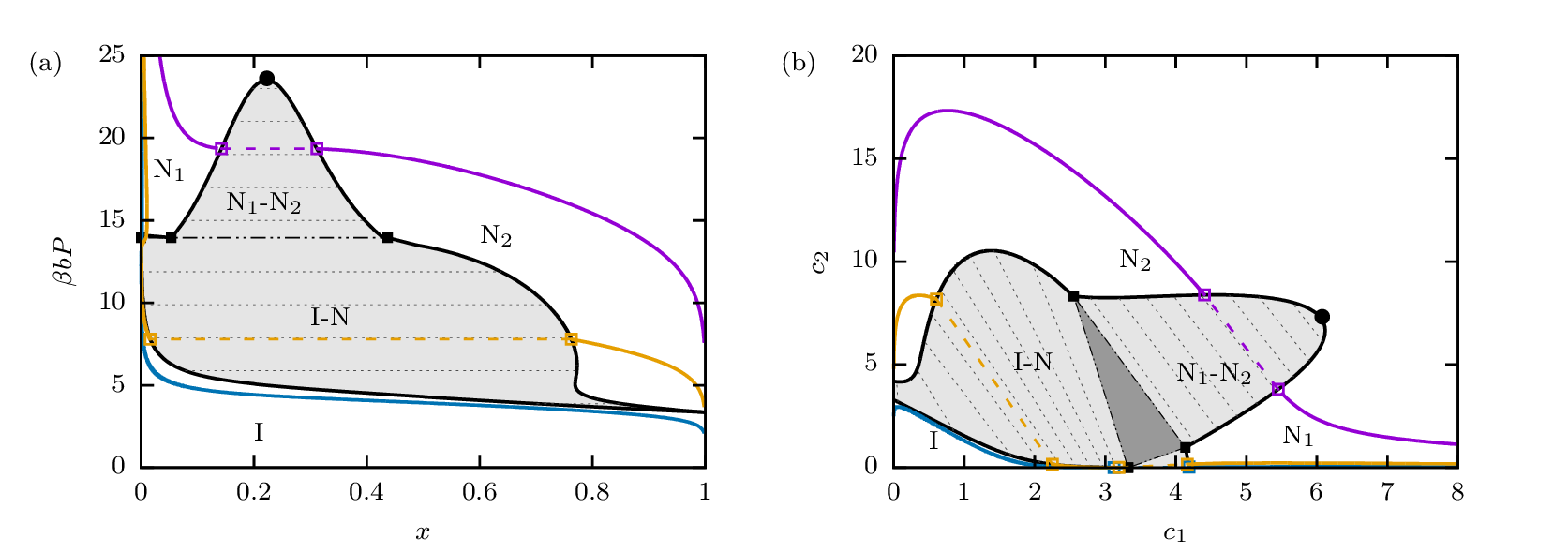}
            \caption{Bulk phase diagram for diameter ratio $d=4.2$ in (a) the composition $x$ - pressure $P$ and (b) the density $c_1$- density $c_2$ representations. The tie-lines connecting coexisting isotropic-nematic (I-N) and nematic-nematic (N$_1$-N$_2$) phases are shown by dashed lines. The nematic-nematic critical point is denoted by a black dot and the isotropic-nematic-nematic triple point by black squares. The colored curves indicate sedimentation paths discussed in Section \ref{sect:sedimentation}.}
            \label{fig:d4p2}
     \end{figure*}

The phase behavior becomes more complicated for the diameter ratio of $d=4.2$. For this diameter ratio, there is not only an isotropic-nematic demixing, but also a nematic-nematic (N$_1$-N$_2$) demixing as shown in Fig.~\ref{fig:d4p2} in (a) in the composition $x$-pressure $P$ and in (b) in the density $c_1$-density $c_2$ representation. Here we denote the nematic phase rich in thin rods by N$_1$ and the nematic phase rich in thick rods by N$_2$. In addition, we have an isotropic-nematic-nematic (I-N$_1$-N$_2$) triple point (black squares) and a nematic-nematic (N$_1$-N$_2$) upper critical point (black dot). We note that while the I-N binodals of Fig.~\ref{fig:d4p2}(a) are in good agreement with results from Ref.~\citen{vanRoij1998}, the N$_1$-N$_2$ critical point occurs at about half of the pressure and at slightly higher $x$ than was found in Ref.~\citen{vanRoij1998}. However, since we have checked and demonstrated in Fig.~\ref{fig:psi} that our $\theta$-grid is sufficiently accurate, we believe this difference is caused by the sensitivity of the phase boundaries to small numerical inaccuracies in the orientation distribution function calculated within the high-pressure scaling analysis.

\begin{figure*}[htbp]
        \centering
                \includegraphics[width=\linewidth]{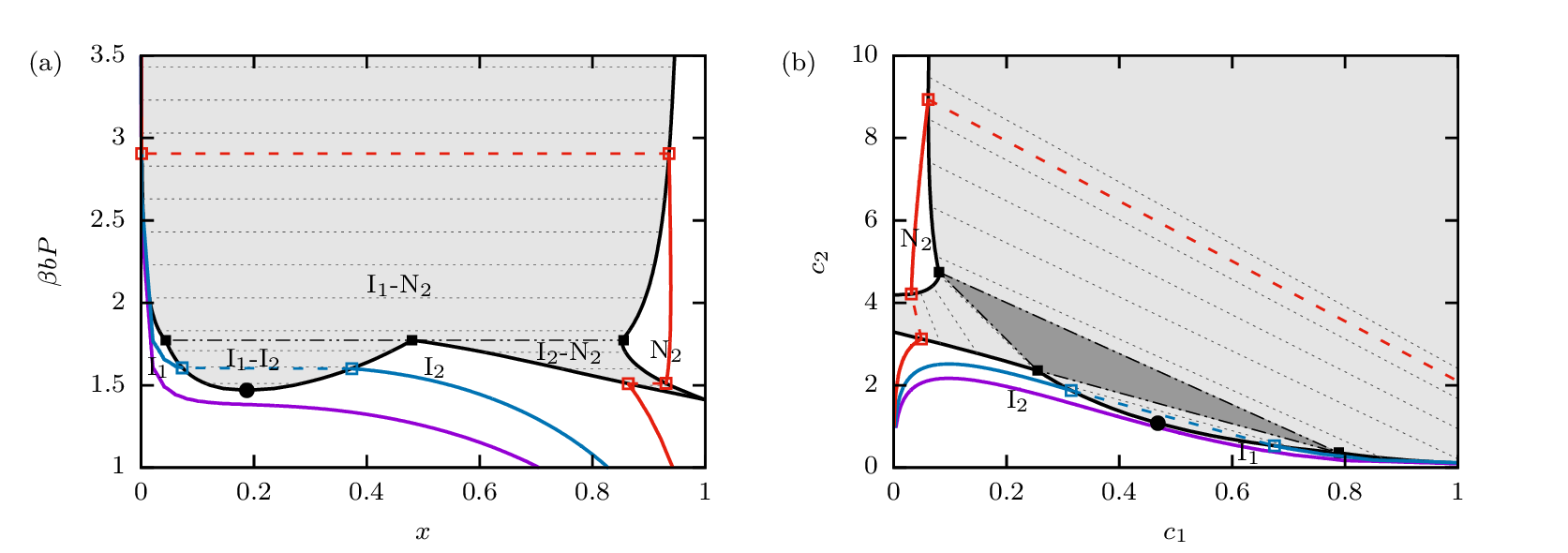}
            \caption{Bulk phase diagrams for diameter ratio $d=10$ in (a) the composition $x$ - pressure $P$ and (b) the density $c_1$- density $c_2$ representations. The tie-lines connecting coexisting isotropic-isotropic (I$_1$-I$_2$), isotropic-nematic (I$_2$-N$_2$ and I$_1$-N$_2$) phases are shown by dashed lines. The isotropic-isotropic critical point is denoted by a black dot and the isotropic-isotropic-nematic triple point by black squares. The colored curves indicate sedimentation paths discussed in Section \ref{sect:sedimentation}.}
            \label{fig:d10}
     \end{figure*}

The final diameter ratio we consider here is $d=10$.  As shown in Fig.~\ref{fig:d10}, there is an isotropic-isotropic (I$_1$-I$_2$) demixing at low pressures ending in a lower I$_1$-I$_2$ critical point (black dot), and an I$_1$-I$_2$-N$_2$ triple point (black squares). In addition, this diameter ratio (and any with $d \geq 4.29$) also features an I-N$_1$-N$_2$ triple point at pressures beyond the scale of Fig.~\ref{fig:d10}, with the nematic-nematic demixing persisting as $P \to \infty$.\cite{vanRoij1998} For both Fig.~\ref{fig:d10}(a) the composition $x$-pressure $P$ and Fig.~\ref{fig:d10}(b) the density $c_1$-density $c_2$ representation the plot ranges are limited to low pressures/ densities for clarity. 

\section{Sedimentation}\label{sect:sedimentation}

We now consider a binary colloidal mixture in sedimentation-diffusion equilibrium, following the theoretical framework but not the notation of Refs.~\citen{Schmidt2004,deLasHeras2012,deLasHeras2013,deLasHeras2015}. In the employed local density approximation (LDA), the height-dependent local chemical potential of species $\sigma=1,2$ in the phase $\alpha=$I,N can be written as
    \begin{equation}\label{eq:LDA}
        \beta \mu_\sigma^\alpha\left( x(z), c(z) \right) =  \beta \mu_\sigma^\text{tot} - z/l_\sigma,
    \end{equation}
where $z$ is the vertical coordinate, $\mu_\sigma^\text{tot}$ is the total chemical potential of species $\sigma$ (which is a spatial constant in equilibrium), and $l_\sigma= k_BT/(m_\sigma g)$ is the gravitational length of species $\sigma$ with $m_\sigma$ the buoyant mass and $g$ the acceleration due to gravity. Eliminating $z$ from Eq.~\eqref{eq:LDA} for $\sigma=1,2$, allows us to write the linear relation
    \begin{equation}\label{eq:sedimentation}
        \beta\mu_2^\alpha(\mu_1^\alpha)=a+s\beta\mu_1^\alpha,
    \end{equation}
where the ``composition'' variable $a$ and the slope $s$ are defined by
    \begin{align}
        a&=\beta\mu_2^\text{tot}-s\beta\mu_1^\text{tot},\nonumber\\
        s&=m_2/m_1.
    \end{align}
Note that a large positive $a$ implies a $2$-rich sample and that $s$ can take positive or negative values depending on the signs of the buoyant masses.

For a given $a$ and $s$, i.e. for a given overall composition of the sample and the buoyant masses of the particles, Eq.~\eqref{eq:sedimentation} gives a sedimentation path through the phase diagram in the $\mu_1$-$\mu_2$ representation. The crossing of a binodal in the $\mu_1$-$\mu_2$ phase diagram corresponds to the interface between two phases in a test tube and we can use Eq.~\eqref{eq:LDA} to relate a difference in heights in the sample $\Delta z = z_1 - z_2$ to a difference in either of the chemical potentials as
    \begin{equation}\label{eq:deltaMu}
      \beta  \Delta \mu_\sigma = \beta\mu_{\sigma}(x(z_1),c(z_1))- \beta\mu_{\sigma}(x(z_2),c(z_2)) = -\Delta z / l_\sigma. 
    \end{equation}
 Since following one line (segment) in the $\mu_1$-$\mu_2$ phase diagram gives the height-dependent sequence of phases in a test tube, all possible lines in this representation give all possible stacking sequences. A stacking sequence is thus determined by the slope $s$ (the ratio of the buoyant masses of the two species), the length of the path (proportional to the height of the container), the composition variable $a$ (determined by the overall composition and concentration), and the direction of the path (determined by the signs of the buoyant masses). \cite{Schmidt2004,deLasHeras2012,deLasHeras2013,deLasHeras2015}

In order to construct a stacking diagram, we must find boundaries between stacking sequences in the $s$-$a$ plane. Following Refs.~\citen{deLasHeras2013,deLasHeras2015} we distinguish three types of features in these diagrams: (1) Sedimentation binodals-- the set of all lines tangent to a binodal in the $\mu_1$-$\mu_2$ phase diagram, (2) terminal lines-- the set of all lines through an end point of a binodal (triple point or critical point), and (3) asymptotic terminal lines-- the set of lines with the asymptotic slope of a bulk binodal that does not terminate at a finite chemical potential. Note that both horizontal and vertical asymptotic terminal lines occur for a binary mixture approaching a pure composition ($\mu_\sigma \to -\infty$ corresponding to $N_\sigma/V \to 0$), however, in the $s$-$a$ representation, the vertical asymptote does not appear as it corresponds to a line with infinite slope $s$. 

\begin{figure*}[htbp]
        \centering
                \includegraphics[width=\linewidth]{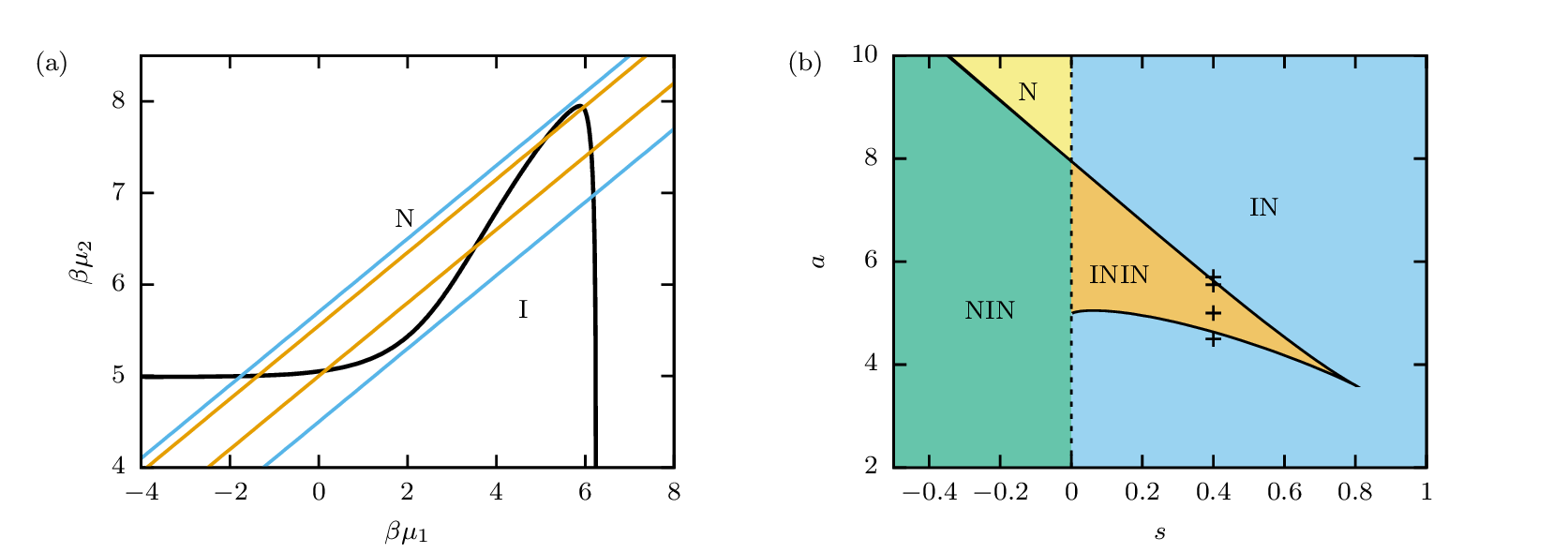}
            \caption{(a) Bulk phase diagram for diameter ratio $d=3.5$ in the plane of the chemical potentials $\mu_1$-$\mu_2$, where lines show sedimentation paths with slope $s=0.4$ and intercept $a$ of $4.5$ (lower blue), $5$ (lower orange), $5.55$ (upper orange) and $5.7$ (upper blue). (b) Stacking diagram for $d=3.5$ in the slope $s$, composition $a$ plane (see text), where the plus symbols represent the four sedimentation lines from (a). See text for explanation of the regions and curves.}
            \label{fig:d3p5b}
     \end{figure*}

We now present the phase diagrams of Section \ref{sect:theory} in the plane of chemical potentials, from which we construct corresponding stacking diagrams. For the diameter ratio of $d=3.5$, the chemical potential $\mu_1$- chemical potential $\mu_2$ phase diagram, shown in Fig.~\ref{fig:d3p5b}(a), consists of a single binodal separating the isotropic phase (I) from the nematic phase (N), with a vertical and a horizontal asymptote corresponding to $x=0,1$. Even this simple phase diagram gives rise to four different stacking sequences in our stacking diagram, as shown in Fig.~\ref{fig:d3p5b}(b). Here we label all stacking sequences from bottom to top of sample under the assumption that $m_1<0$; if we had taken the assumption instead that $m_1>0$, the labels would describe the path from top to bottom of test tube. There are two stacking sequence boundaries in Fig.~\ref{fig:d3p5b}(b): one sedimentation binodal (solid line) and one asymptotic terminal line (dotted line). The most complicated of four sequences, ININ, is made possible by the existence of an inflection point in the binodal in Fig.~\ref{fig:d3p5b}(a), which allows the binodal in the $\mu_1$-$\mu_2$ representation to be crossed three times by a straight line. We have added lines to Fig.~\ref{fig:d3p5b}(a), which show four possible sedimentation paths corresponding to distinct points in Fig.~\ref{fig:d3p5b}(b), with the colors of the sedimentation path corresponding to those of its stacking sequence in Fig.~\ref{fig:d3p5b}(b) (crosses are shown here for each sedimentation path in (a)). For the four different fixed $s$ and $a$, we have sedimentation lines in Fig.~\ref{fig:d3p5b}(a), which can be transformed to sedimentation contours in the phase diagrams of Fig.~\ref{fig:d3p5}(a) and (b). These contours must satisfy  $a = \beta\mu_2^\alpha(c,x) - s \beta\mu_1^\alpha(c,x)$ from which the contours in the $c_1$-$c_2$ representation follow readily. Moreover, since we know $P(c,x)$, by inverting this to find $c(P,x)$ we can plot contours in the ($x$, $P$) plane as well. 

\begin{figure*}[htbp]
        \centering
                \includegraphics[width=\linewidth]{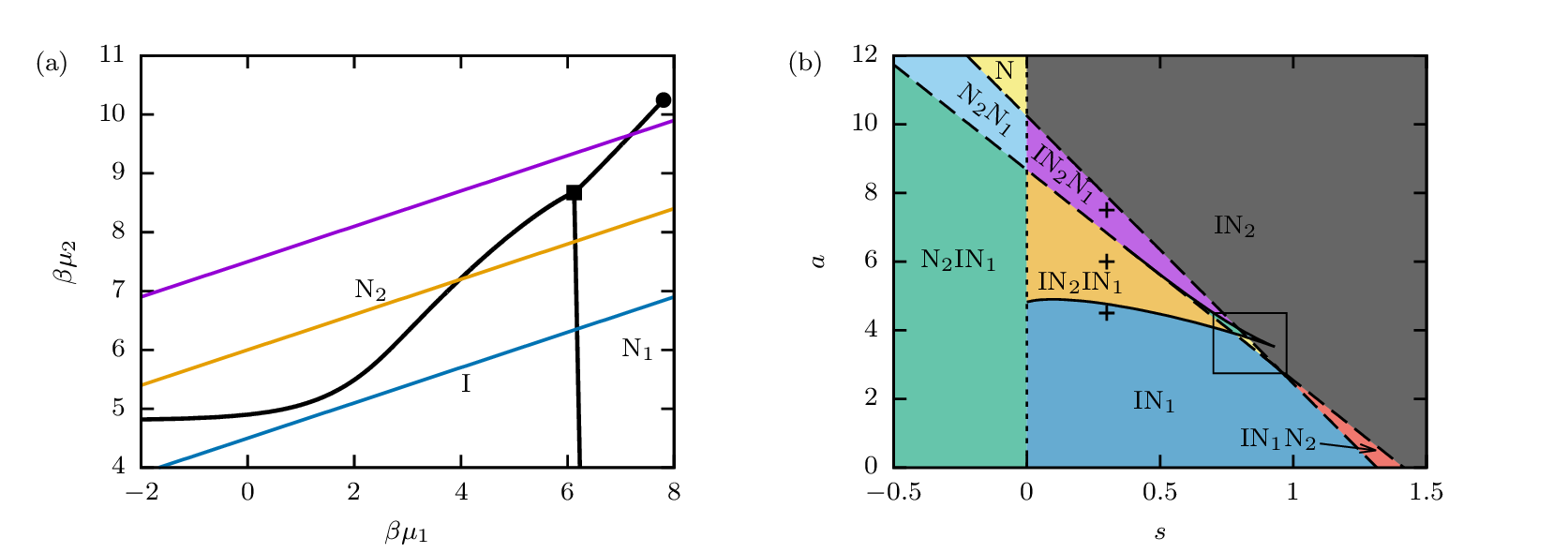}

                \includegraphics[width=.5\linewidth]{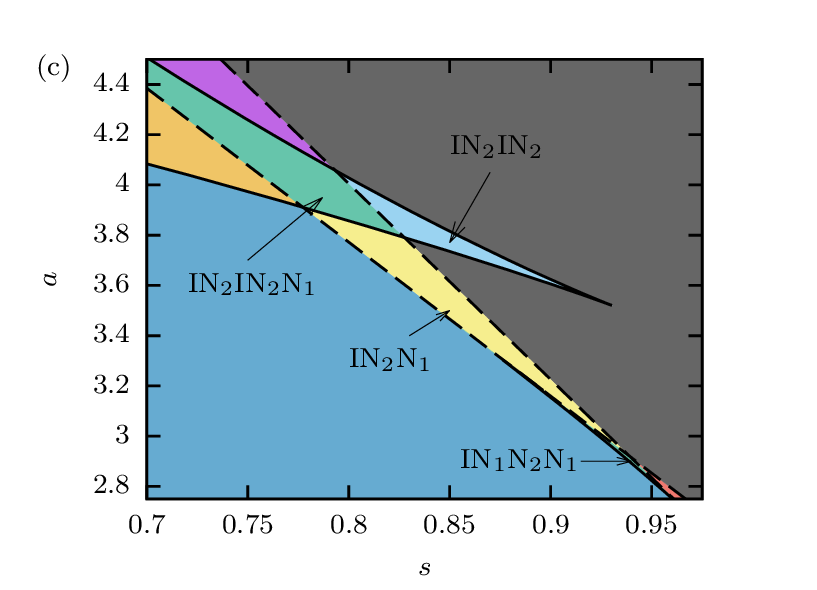}
            \caption{(a) Bulk phase diagram for diameter ratio $d=4.2$ in the plane of the chemical potentials $\mu_1$-$\mu_2$, where lines show sedimentation paths with slope $s=0.3$ and intercept $a$ of $4.5$ (blue), $6$ (yellow), and $7.5$ (purple). (b) Stacking diagram for $d=4.2$ in the slope $s$, composition $a$ plane (see text), where the plus symbols represent the sedimentation lines from (a). In (c) we show a zoomed in version of (b). See text for explanation of the regions and curves.}
            \label{fig:d4p2b}
     \end{figure*}

For the diameter ratio of $d=4.2$, we show the $\mu_1$-$\mu_2$ representation in Fig.~\ref{fig:d4p2b}(a). As discussed in the previous section, for this diameter ratio there is not only isotropic-nematic demixing, but also nematic-nematic (N$_1$-N$_2$) demixing. As in Fig.~\ref{fig:d4p2}, we have a I-N$_1$-N$_2$ triple point (black square) and a N$_1$-N$_2$ critical point (black dot). In Fig.~\ref{fig:d4p2b}(b) we show the stacking diagram in the $s,a$ plane, and in Fig.~\ref{fig:d4p2b}(c) we show a zoomed-in version of this. The stacking diagram has five different boundaries: two sedimentation binodals (solid lines), one asymptotic terminal line (dotted line), and two terminal lines corresponding to the triple point and the critical point (dashed lines). Altogether, we find twelve regions, eleven of which are distinct stacking sequences (note that some colors are repeated), some of which are so small that they can only be seen in the zoom-in shown in (c). The smallest of these, IN$_1$N$_2$N$_1$ in the lower right of (c), comes from the fact that the calculated N$_1$-N$_1$ binodal in (a) is very close to, but not perfectly linear; however, these tiny regions may not be easily experimentally accessible, so here we focus on a few of the larger regions. Three of the larger regions are illustrated by the sedimentation lines shown in Fig.~\ref{fig:d4p2b}(a), corresponding to the crosses in Fig.~\ref{fig:d4p2b}(b), with their equivalent sedimentation contours shown in the $x$-$P$ and $c_1$-$c_2$ representations of the phase diagram in the previous section (Fig.~\ref{fig:d4p2}). Clearly, the crossing of a $\mu_1$-$\mu_2$ line can correspond to a huge density or composition jump in the $x$-$P$ and $c_1$-$c_2$ representations.

\begin{figure*}[htbp]
        \centering
                \includegraphics[width=\linewidth]{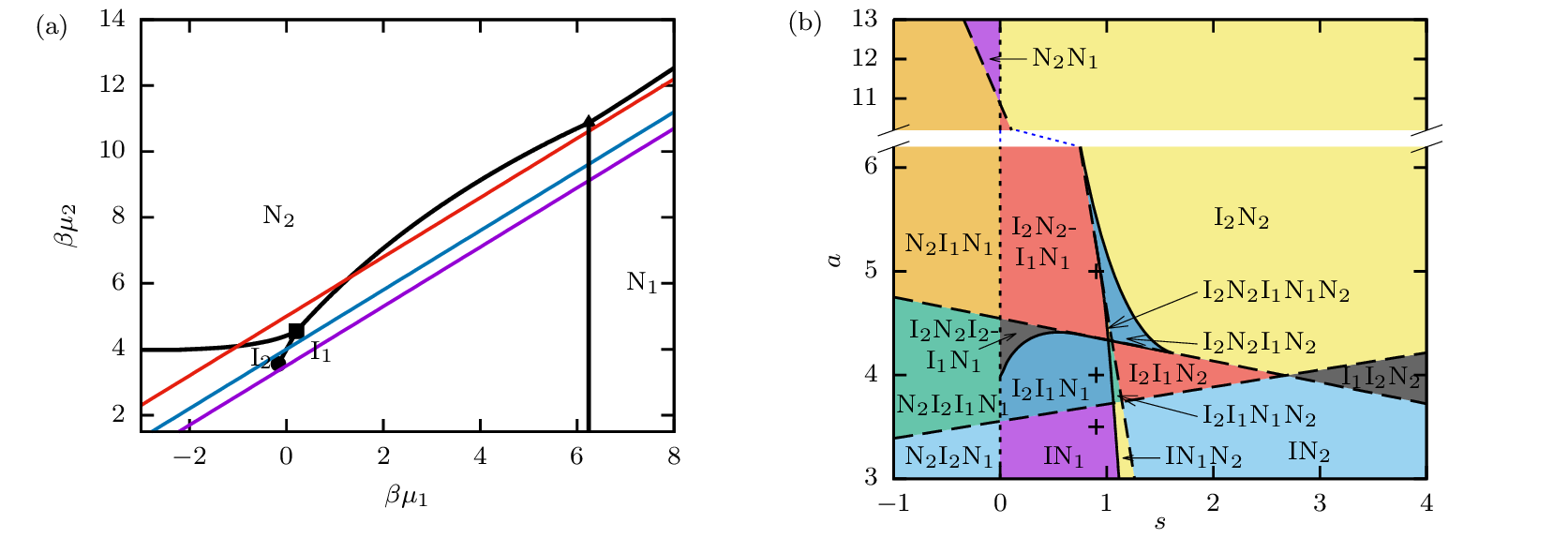}
            \caption{(a) Bulk phase diagram for diameter ratio $d=10$ in the plane of the chemical potentials $\mu_1$-$\mu_2$, where lines show sedimentation paths with slope $s=0.9$ and intercept $a$ of $3.5$ (purple), $4$ (blue), and $5$ (red). (b) Stacking diagram for $d=10$ in the slope $s$, composition $a$ plane (see text), where the plus symbols represent the sedimentation lines from (a). See text for explanation of the regions and curves.}
            \label{fig:d10b}
     \end{figure*}

The final diameter ratio we consider is $d=10$ (Fig.~\ref{fig:d10b}). As shown in the plane of chemical potentials in Fig.~\ref{fig:d10b}(a), we have an I-N$_1$-N$_2$ triple point (black triangle), an I$_1$-I$_2$ critical point (black dot), and an I$_1$-I$_2$-N$_2$ triple point (black square). The N$_1$-N$_2$ binodal has a well-defined slope as $\mu_{1,2} \to \infty$.\cite{deLasHeras2015} In Fig.~\ref{fig:d10b}(a) we also show three examples of sedimentation paths corresponding to three points in the stacking diagram shown in (b) and three sedimentation contours in the $x$-$P$  and $c_1$-$c_2$ phase diagrams of Fig.~\ref{fig:d10}. In Fig.~\ref{fig:d10b}(b), we see that the stacking diagram is extremely rich; there are three sedimentation binodals (solid lines), one asymptotic terminal line (dotted line), and three terminal lines corresponding to the two triple points and one critical point (dashed lines). Altogether, we find sixteen distinct stacking sequences (though colors are repeated in Fig.~\ref{fig:d10b}(b)).  

\begin{figure*}[htbp]
        \centering
                \includegraphics[width=\linewidth]{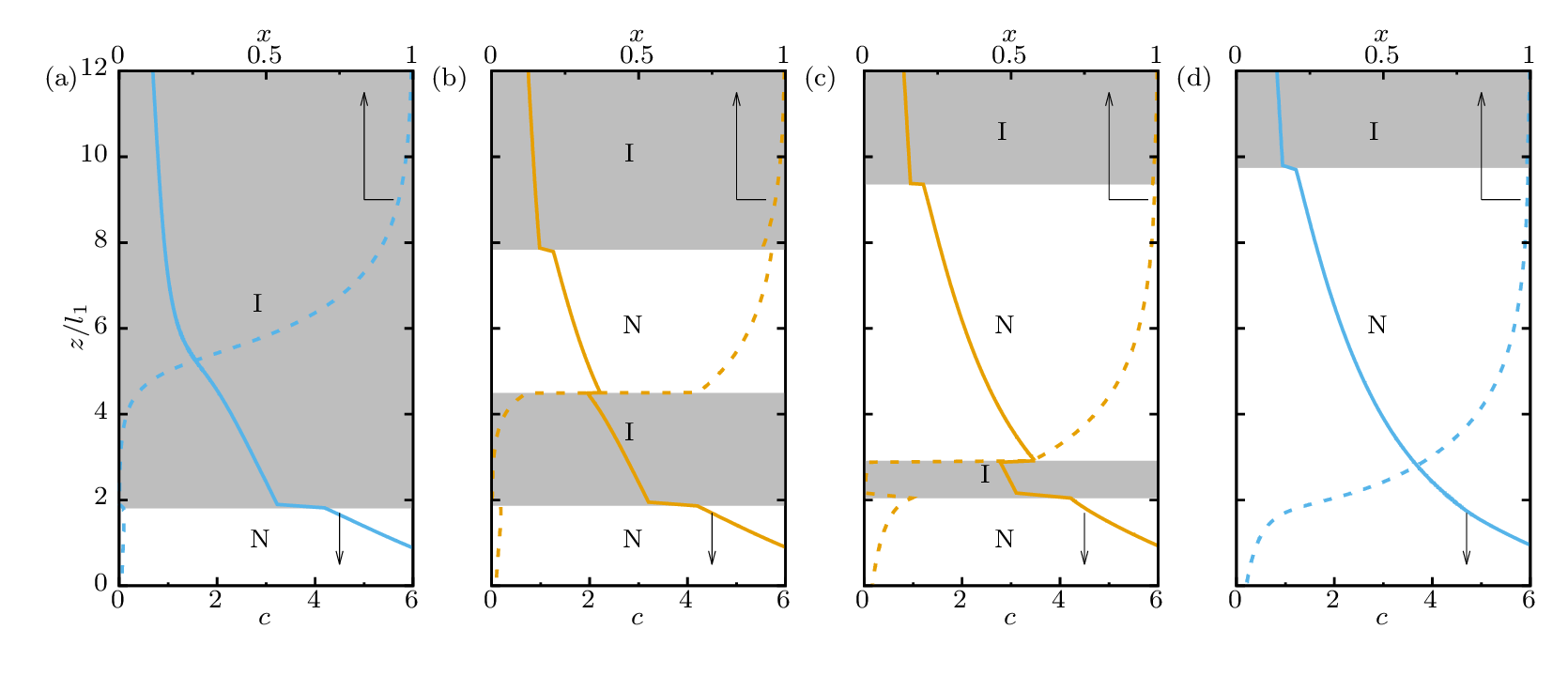}
            \caption{Composition $x$ (dashed) and density $c$ (solid) as a function of height in test tube $z$ (rotated so $z$ is on vertical axis) for diameter ratio $d=3.5$, slope $s=0.4$, and composition variable (a) $a=4.5$, (b) $a=5$, (c) $a=5.55$, and (d) $a=5.7$ (see also Figs.~\ref{fig:d3p5} and \ref{fig:d3p5b}). Gray indicates an isotropic phase (I) while the white indicates a nematic phase (N) and the arrows are to emphasize that the dashed curves correspond to the upper x-axes while the solid curves correspond to the lower x-axes. Here we assume both species have positive buoyant masses.}
            \label{fig:d3p5c}
     \end{figure*}

We use the fact that a difference in heights is proportional to a difference in chemical potential (see Eq.~\eqref{eq:deltaMu}), to plot the dependence of composition $x$ and density $c$ on height $z$ (in units of the gravitational length $l_1$ of species one). For diameter ratio $d=3.5$, we show in Fig.~\ref{fig:d3p5c} four ``test tubes'' corresponding to the four sedimentation contours with $s=0.4$ and (a) $a=4.5$, (b) $a=5$, (c) $a=5.55$, and (d) $a=5.7$ displayed in Figs.~\ref{fig:d3p5} and \ref{fig:d3p5b}. Here we take both buoyant masses to be positive (which leads to inverted stacking sequences with respect to previous figures where we assumed $m_1<0$). We arbitrarily choose a test tube of height $12 l_1$ by limiting $\beta\mu_1 \in [-4,8]$ in all four cases (which also limits $\mu_2$ for a given $a$ and $s$); a shorter test tube could limit the number of layers that are seen. In Fig.~\ref{fig:d3p5c}(a) we see a stacking sequence with a thin-rich nematic phase on the bottom and above that an isotropic phase that transitions from thin-rich to thick-rich with increasing height. In Fig.~\ref{fig:d3p5c}(b) there are four sedimented layers, from bottom to top: a thin-rich nematic, a thin-rich isotropic, a thick-rich nematic, and a thick-rich isotropic. Here we see that the density and composition are non-monotonic in $z$. Similarly, there are four layers in Fig.~\ref{fig:d3p5c}(c), with the bottommost isotropic layer becoming thinner as the middle nematic layer grows thicker, and the changes in composition and density at the interfaces between phases being more pronounced than in (b). Finally, in Fig.~\ref{fig:d3p5c}(d), there are once again only two layers, namely a large nematic layer which transitions from thin- to thick-rich with a thick-rich isotropic floating on top.

\section{Summary and Discussion}\label{sect:concl}
We studied the sedimentation of binary mixtures of thick and thin hard rods, focusing on diameter ratios $d=3.5, 4.2, 10$ by considering sedimentation paths as straight lines in the plane of chemical potentials. This method is based on the local density approximation, which is known to be good for most colloidal rod systems, since the gravitational lengths of these systems of roughly a centimeter are much larger than all relevant correlations lengths. By considering all sedimentation paths, we constructed stacking diagrams. Experimentally, it is possible to sample the different regions of the stacking diagram by changing the composition of the mixture, the height of the container, and the maximum accessible density by e.g. (ultra)centrifugation.\cite{deLasHeras2013,deLasHeras2015,Svedberg1940,Rasa2004,Planken2010}

We found that even the simplest phase diagram (for diameter ratio $d=3.5$), with only spindle-shaped isotropic-nematic demixing, led to a stacking diagram with four regions, including sequences with an isotropic phase floating between two nematic phases. Adding a single binodal and associated critical and triple points, as occurs for diameter ratio $d=4.2$, increases this number of regions to eleven distinct stacking sequences. Our richest phase diagram (for diameter ratio $d=10$),  contained two triple points and one critical point, produced a stacking diagram with sixteen distinct regions, including two with five sedimented layers. In addition to this straightforward application of the method developed in Refs.~\citen{deLasHeras2013,deLasHeras2015}, we also translated sedimentation lines from the plane of chemical potentials to paths in the more familiar representations of composition-pressure and density-density, and showed the density and composition profiles as a function of height for the diameter ratio $d=3.5$.

The sedimentation contours shown in Figs.~\ref{fig:d3p5}-\ref{fig:d10} are very ``jumpy'', with several possibly surprising effects of gravity on binary mixtures present. In Fig.~\ref{fig:d3p5c}, we see for instance that there are rapid changes in composition and density as a function of height in a ``test tube'', which change non-monotonically along the sedimentation path. Figures~\ref{fig:d3p5c}(b) and (c) show the stacking sequences NINI, which is possible even though there is no triple point in the bulk phase diagram. This leads us to the point that studying the sedimentation of mixtures experimentally demands great care in drawing conclusions about the bulk phase diagrams. If the concentrations of the two coexisting phases are measured sufficiently close to the interface between two layers, it should indeed be possible to experimentally reconstruct the bulk phase diagram on the basis of measurements of many sample compositions. Suppose however, that a layer is very thin (e.g. even thinner than the middle isotropic phase in Fig.~\ref{fig:d3p5c}(c)) such that it is experimentally invisible or missed. The observed sedimentation sequence might then be assumed to imply that a thick-rich nematic coexists with a thin-rich nematic. This potential ``mistake'' is illustrate by the red dashed lines in Figs.~\ref{fig:d3p5}(a) and (b) that show a hypothetical ``false'' nematic-nematic coexistence line, even though there is no nematic-nematic demixing present in the bulk phase diagram.

In Ref.~\citen{Purdy2005}, the phase behavior of binary mixtures of fd virus and fd coated with neutral polymer polyethylene glycol (PEG) was studied. The diameter of the bare fd was varied by changing the ionic strength, which varied its effective diameter and thus the diameter ratio, while the buoyant mass ratio remained fixed at $0.3 \leq s \leq 0.7$ and the gravitational length of the bare fd was approximately $4$ cm. For diameter ratios $d\geq 3$ the authors find N$_\text{thin}$-I$_\text{thin}$-N$_\text{thick}$ and N$_\text{thin}$-N$_\text{thick}$ stacking sequences. We show in Figs.~\ref{fig:d3p5} and \ref{fig:d3p5b} that an NIN stacking sequence can occur without the presence of triple point, and as argued above, an observed coexistence between two nematics could possibly be a NIN stacking sequence with a very thin isotropic layer. In the case of Ref.~\citen{Purdy2005} a convincing phase diagram which included an N$_\text{thin}$-I$_\text{thin}$-N$_\text{thick}$ triple point and nematic-nematic demixing was presented based on many measurements, and was also qualitatively backed up by theoretical calculations.\cite{Varga2005} However, we do wish to stress that a single observation of a stacking sequence of the type ABA of phases A and B in a binary mixture does not imply the existence of an AAB triple point in the bulk phase diagram.

Though a single sedimentation density profile for a one-component system yields the full equation of state of the system, for colloidal mixtures this is clearly not so. Here we illustrated the inherent difficulties in drawing conclusions about bulk phase diagrams of binary mixtures. So although systematic measurements of SD properties do allow for conclusions to be drawn about bulk phase diagrams of mixtures, one should be wary and study a large set of thermodynamic state points.

Onsager theory only gives quantitative results for needle-like, rigid rods but it can be readily generalized to study more realistic systems as well. Both including finite size effects\cite{Varga2005} and adding flexibility\cite{Dennison2011a,Dennison2011b} were shown to give better agreement with experimental results for binary mixtures of fd virus. However, the ungeneralized Onsager theory did capture qualitative features of the phase behavior including the isotropic-nematic, nematic-nematic, and isotropic-isotropic demixing.

Here we only considered homogeneous phases, but of course at higher pressures one should expect phases with partial positional ordering, such as smectic phases. For a one-component system of needle-like rods, the nematic-smectic transition occurs at pressures far beyond the isotropic-nematic transition, however, for rods with a small aspect ratio $L/D \sim 4 -5$ the nematic regime is small and direct isotropic-smectic transitions are to be expected.\cite{Bolhuis1997} For a binary mixture, these inhomogeneous phases will also lead to richer phase diagrams and hence richer stacking diagrams. For shorter rods, smectic phases can preempt isotropic-nematic and nematic-nematic transitions, thus significantly altering the phase diagram.\cite{Cinacchi2006}

We note that three-component systems, let alone polydisperse ones, are expected to be even richer, but also considerably more complicated to analyze in full detail. This is left for future studies.

\section*{Acknowledgments}

We thank Marjolein Dijkstra for helpful discussions. This work is part of the D-ITP consortium, a program of the Netherlands Organization for Scientific Research (NWO) that is funded by the Dutch Ministry of Education, Culture and Science (OCW). We also acknowledge financial support from an NWO-VICI grant.

\bibliography{binaryMixtures}
\bibliographystyle{rsc} 

\end{document}